\documentclass[12pt]{article}
\usepackage{amsmath}
\usepackage{amsfonts}
\usepackage{latexsym}
\usepackage{bm}

\advance \textheight by \topskip
\advance \textheight by 1pt
\makeatother
\def\bea{\begin{eqnarray}}
\def\ena{\end{eqnarray}}
\def\non{\nonumber}
\newcommand{\qed}{\hbox{\rule[-2pt]{3pt}{6pt}}}
\def\sgn{{\rm sgn}}
\def\Pf{{\rm Pf}}

\newtheorem{prop}{Proposition}
\newtheorem{theorem}{Theorem}

\newtheorem{lemma}{Lemma}
\newtheorem{cor}{Corollary}

\newtheorem{ex}{Example}

\def\pf{{\it Proof.}\,}

\def\qed{$\Box$}

\title{On the Expansion Coefficients  of Tau-function of the BKP Hierarchy}

\author{
Yoko Shigyo\thanks{
e-mail: yoko.shigyo@gmail.com}\\
Department of Mathematics, Tsuda College,\\
Kodaira, Tokyo, 187-8577, Japan \\
\\
\\
\\
 \\
}
\date{}
\begin{document}
\maketitle

\begin{abstract}
We study the series expansion of the tau function of the BKP hierarchy applying the addition formulae of the BKP hierarchy.
Any formal power series can be expanded in terms of Schur functions.
It is known that, under the condition $\tau(x)\neq0$, a formal power series $\tau(x)$ is a solution of the KP hierarchy if and only if its coefficients of Schur function expansion are given by the so called Giambelli type formula.
A similar result is known for the BKP hierarchy with respect to Schur's Q-function expansion under a similar condition.
In this paper we generalize this result to the case of $\tau(0)=0$.
\end{abstract}

\section{Introduction} 
The tau function of the KP hierarchy can be expanded in terms of Schur functions $\chi_\mu(x)$, $x=(x_1,x_2,\cdots)$ as
\bea
&&
\tau(x)=\sum_{\mu}\xi_{\mu}\chi_{\mu}(x),
\label{eq:1}
\ena
where $\mu$ runs over all partitions.
If $\tau(0)=1$, $\xi_\mu$ for an arbitrary partition $\mu$ is written in terms of $\xi_{(i|j)}$ corresponding to the hook diagram $(i|j)$ as
\bea
&&
\xi_{(i_1\cdots,i_k|j_1,\cdots,j_k)}=\det(\xi_{(i_r|j_s)})_{1\le r,s\le k},
\label{eq:2}
\ena
where $(i_1\cdots,i_k|j_1,\cdots,j_k)$ is the Frobenius' notation for a partition \cite{Mac}.
This formula is called the Giambelli formula since it can be considered as the interpretation of the Giambelli formula for Schur function \cite{Mac} to $\xi_\mu$.
It is known that, under the condition $\tau(0)=1$, (\ref{eq:1}) is a solution of the KP hierarchy if and only if (\ref{eq:2}) holds for any partition $(i_1\cdots,i_k|j_1,\cdots,j_k)$.
This follows from Sato's theory of the KP hierarchy \cite{SS1, EH1, SN1, NT1}.

A formal power series $\tau(x)$, $x=(x_1,x_3,\cdots)$ can be expanded in terms of Schur's Q-function as
\bea
&&
\tau(x)=\sum_{\mu}\xi_{\mu}Q_{\mu}\left(\frac{x}{2}\right),
\label{eq:3}
\ena
where $\mu$ runs over all strict partitions.
In this case, if $\tau(0)=1$, then $\xi_\mu$ for a strict partition $\mu=(\mu_1,\cdots,\mu_{2l})$ satisfy
\bea
&&
\xi_{(\mu_1,\cdots,\mu_{2l})}=\Pf(\xi_{(\mu_i,\mu_j)}),
\label{eq:4}
\ena
where Pf $(a_{ij})$ denotes the Pfaffian of a skew symmetric matrix $(a_{ij})$ (see (\ref{**}) for more precise definition and notation).
Similarly to the KP case, under the condition $\tau(0)=1$, (\ref{eq:3}) is a solution of the BKP hierarchy if and only if (\ref{eq:4}) holds for any strict partition $\mu$.
This is proved in \cite{DJKM2} using the free fermion construction of the tau function.

There are many solutions for which $\tau(0)=0$.
To the best of the author's knowledge the formula corresponding to (\ref{eq:2}) or (\ref{eq:4}) in this case is not known in general.
In this paper we have found a formula of the form (\ref{eq:4}) in the case $\tau(0)=0$, which we call the Giambelli type formula, and have proved that (\ref{eq:3}) is a solution to the BKP hierarchy if and only if the Giambelli type formulae hold.
It seems that the KP case is more involved and is now under investigation. 

Let us describe our results more precisely.
For a strict partition $\lambda=(\lambda_1,\cdots,\lambda_M)$ of length $M$ we consider a function of the form
\bea
&&
\tau(x)=Q_{\lambda}(\frac{x}{2})+\sum_{|\mu|>|\lambda|}\xi_{\mu}Q_{\mu}(\frac{x}{2}).
\label{eq:5}
\ena
In order to give a formula for $\xi_\mu$ we introduce notation for Pfaffian.
For symbols $X_1,\cdots,X_N$, $N\in\mathbb{N}$ with $X_i=\Lambda,\Lambda^{(j)}\,(1\le j\le M),n,\,n\in\mathbb{N}$, we denote by $(X_i,X_j)(=-(X_j,X_i))$ the $(i,j)$ component of a skew symmetric matrix and by $(X_{i_1},\cdots,X_{i_{2m}})$ the Pfaffian of the matrix $((X_{i_r},X_{i_s}))_{r,s}$ (See Example 1).
We define
\bea
&&
(\Lambda^{(i)},n)=\xi_{(\lambda_1,\cdots,\hat{\lambda}_i,\cdots,\lambda_M,n)},
\non
\\
&&
(\Lambda,n)=\xi_{(\lambda_1,\cdots,\lambda_M,n)},
\non
\\
&&
(n_i,n_j)=\xi_{(\lambda_1,\cdots,\lambda_M,n_i,n_j)},
\non
\\
&&
(\Lambda,\Lambda^{(i)})=(\Lambda^{(i)},\Lambda^{(j)})=0.
\non
\ena
Then we prove that (\ref{eq:5}) is a solution to the BKP hierarchy if and only if $\xi_\mu$ for $\mu=(\mu_1,\cdots,\mu_l)$ being a partition of length $l$ is given by the following formula:
\bea
&&
\xi_{\mu}=(\Lambda^i,\Lambda^{(1)},\cdots,\Lambda^{(M)},\mu_1,\cdots,\mu_{l}),
\label{eq:6}
\ena
where $i=0,1$ is chosen in such a way that $i+M+l$ is even and $\Lambda^0$ means that $\Lambda$ is not inserted there.
In the case of $\lambda=\emptyset$ this formula recovers the result of the case of $\tau(0)=1$.

This result is proved using addition formulae for the tau function of the BKP hierarchy.
In our previous paper \cite{S1} we have studied the addition formulae of the BKP hierarchy and proved that the simplest case of the addition formula is equivalent to the BKP hierarchy itself.
In this sense the addition formulae have enough information on the BKP hierarchy.

We derive the relations among $\{\xi_\mu\}$ by expanding addition formula.
By solving these relations we prove (\ref{eq:6}).
It should be mentioned that it is difficult to derive (\ref{eq:6}) from the DJKM equation (\ref{3210}), which is used to derive (\ref{eq:4}).

Recently Giambelli and Jacobi-Trudi type formulae for the expansion coefficients attract much attention in relation to the study of 2 dimensional solvable lattice models \cite{AKLTZ, NZ1, KN1} and higher genus theta functions \cite{EH1,N1}. 
It is interesting to study relations of our results to such subjects.

Finally, it is known that totality of solutions to the BKP hierarchy is parametrized by the infinite dimensional orthogonal Grassmann manifold \cite{DJKM1,DJKM2}.
It should be clarified whether our result can be proved from this point of view.
It is also interesting to study the relation with cluster algebras \cite{R1}.

This paper consists of two sections and two appendixes.
In section 2, we consider the addition formulae in the BKP hierarchy.
We first review the BKP hierarchy and introduce Schur's Q-function.
Then we study the expansions of the addition formulae in \cite{S1}.
In section 3, we state and prove the main result of this paper.
In appendix A, necessary facts on free fermion and the boson-fermion correspondence are reviewed.
We prove that a formal power series of variable $x=(x_1,x_3,\cdots)$ can be expanded in terms of Schur's Q-function in appendix B.

\section{The BKP hierarchy and its addition formulae}
\subsection{Addition formulae of the BKP hierarchy}
In this section, we review the BKP hierarchy and its addition formulae.

Set
\bea 
&&
[\alpha]_o=(\alpha,\frac{\alpha^3}{3},\frac{\alpha^5}{5},\cdots),\,\,\tilde{\xi}(x,k)=\sum_{n=1}^{\infty}x_{2n-1}k^{2n-1},\,\, x=(x_1,x_3,x_5,\cdots),\,\,y=(y_1,y_3,y_5,\cdots).
\non
\ena
The BKP hierarchy \cite{DJKM1} is a system of non-linear equations for $\tau(x)$ given by
\bea
&&
\oint e^{-2\tilde{\xi}(y,k)}\tau(x-y-2[k^{-1}]_o)\tau(x+y+2[k^{-1}]_o)\frac{dk}{2\pi i k}=\tau(x-y)\tau(x+y),
\non
\ena
where the integral means taking the coefficient of $k^{-1}$ in the expansion of the integrand in the series of $k$.
Let $\alpha_1,\cdots,\alpha_n$ be parameters.
Set $y=\sum_{l=1}^n[\alpha_l]$ and compute the integral by taking residues, then we have the addition formulae for $\tau(x)$ of the BKP hierarchy \cite{S1}.
There are two kinds of addition formulae according as $n$ is odd or even as follows.
If $n$ is odd and greater than or equal to 3, then we have
\bea 
&&
A_{1\dots n}\tau(x)\tau(x+2\sum_{l=1}^{n}[\alpha_l]_o)
\non
\\
&&
=\sum_{i=1}^{n}(-1)^{i-1}\tau(x+2[\alpha_i]_o)A_{1\cdots\hat{i}\cdots n}\tau(x+2\sum_{l=1,l\neq i}^n[\alpha_l]_o).
\label{211}
\ena
If $n$ is even and greater than or equal to 4, then we have
\bea
&&
A_{1\dots n}\tau(x)\tau(x+2\sum_{l=1}^{n}[\alpha_l]_o)
\non
\\
&&
=\sum_{i=1}^{n-1}(-1)^{i-1}\frac{\alpha_{i,n}}{\tilde {\alpha}_{i,n}} \tau(x+2[\alpha_i]_o+2[\alpha_n]_o)A_{1\cdots\hat{i}\cdots n-1}\tau(x+2\sum_{l=1,l\neq i}^{n-1}[\alpha_l]_o).
\label{212}
\ena
Here $A_{1\dots n}$ is defined by
\bea 
&&
A_{1\dots n}=\prod_{i<j}^{n} \frac{\alpha_{ij}}{\tilde{\alpha}_{ij}},\hskip5mm \tilde{\alpha}_{ij}=\alpha_i+\alpha_j,\hskip5mm \alpha_{ij}=\alpha_i-\alpha_j,
\non
\ena
and $\hat{i}$ means that $i$ is removed.

\vskip5mm
\noindent{\bf{Remark.}}\,\,\,
Notice that the coefficient $A_{1\dots n}$ in this paper is defined by an inverse of $A_{1\dots n}$ in \cite{S1}.

\vskip5mm
The following theorem has been proved in \cite{S1}.

\begin{theorem}
The addition formulae (\ref{211}) and (\ref{212}) are equivalent to the BKP hierarchy respectively.
\end{theorem}

\noindent{\bf{Remark.}}\,\,\, 
In \cite{S1} we had proved that the case of $n=3$ of (\ref{211}) is equivalent to the BKP hierarchy instead of the whole addition formulae.

\subsection{Addition formulae in terms of the expansion coefficients}
First we introduce notation for partitions \cite{Mac}.
A partition $\lambda=(\lambda_1,\cdots,\lambda_n)$ is a non-increasing sequence of non-negative integers.
The non-zero $\lambda_i$ are called parts of $\lambda$.
We define the length $l(\lambda)$ of $\lambda$ as the number of  parts of $\lambda$ and the weight $|\lambda|$ of $\lambda$ as the sum of parts of $\lambda$.
We say $\lambda=(\lambda_1,\cdots,\lambda_{2l})$ is a strict partition if $\lambda_1>\cdots>\lambda_{2n}\ge0$.
(This definition of a strict partition is different from that in \cite{Mac}.)
We identify $(\lambda_1,\cdots,\lambda_{2l-1},0)$ and $(\lambda_1,\cdots,\lambda_{2l-1})$.
If $l=0$ then $\lambda$ is considered as $\emptyset$.

To study the series expansion of the tau function of the BKP hierarchy we first introduce Schur's Q-function.
Schur's Q- function is defined for strict partitions.
  
For a non-negative integer $r$ we define the symmetric polynomial $q_r$ of $\alpha=(\alpha_1,\cdots,\alpha_N)$ by
\bea
&&
\sum_{r\ge 0}q_r t^r=\prod_{i=1}^N\frac{1+t\alpha_i}{1-t\alpha_i}.
\non
\ena
For $r>s\ge 0$, we set 
\bea
&&
Q^{sym}_{(r,s)}(\alpha)=q_r q_s +2\sum_{i=1}^s (-1)^i q_{r+i}q_{s-i}.
\non
\ena
If $r<s$, we define $Q^{sym}_{(r,s)}(\alpha)$ as
\bea
&&
Q^{sym}_{(r,s)}(\alpha)=-Q^{sym}_{(s,r)}(\alpha).
\non
\ena
For any strict partitions $\lambda=(\lambda_1,\cdots,\lambda_l)$, Schur's Q-function is defined by 
\bea
&&
Q^{sym}_{\lambda}(\alpha)=\Pf\left(Q^{sym}_{(\lambda_i,\lambda_j)}(\alpha)\right)_{1\le i,j\le 2n},
\non
\ena
where $\Pf(a_{ij})_{1\le i,j\le 2n}$ denotes the Pfaffian of A$=(a_{ij})_{1\le i,j\le 2n}$ (see (\ref{**}) for the definition of Pfaffian in detail).
We set $x_i=(\alpha^i_1+\cdots+\alpha^i_N)/i,\,\,N\ge |\lambda|$.
It is known that $Q^{sym}_{\lambda}(\alpha)$ can uniquely be expressed as a polynomial of $x=(x_1,x_3,x_5,\cdots)$. 
We denote this polynomial by $Q_{\lambda}(x)$.
Then we have the relation
\bea
&&
Q^{sym}_{\lambda}(\alpha)=Q_{\lambda}(x),\,\,\,x_i=\frac{\alpha^i_1+\cdots+\alpha^i_N}{i}.
\non
\ena
We extend the definition of $Q_\lambda$ for an arbitrary permutation $\lambda$ of a strict partition by skew symmetry.
It is possible to expand any formal power series $\tau(x)$ of $x=(x_1,x_3,x_5,\cdots)$ as follows \cite{Y1} (see appendix B):
\bea
&&
\tau(x)=\sum_{\mu}\xi_{\mu}Q_{\mu}(\frac{x}{2}),
\label{eq:51}
\\
&&
\xi_{\mu}=\left.2^{-l(\mu)}Q_{\mu}(\tilde{\partial})\tau(x)\right|_{x=0},\,\,\tilde{\partial}=(\tilde{\partial}_1,\tilde{\partial}_3,\cdots),\,\,\tilde{\partial}_i=\frac{\partial_{x_i}}{i}
\label{eq:111}
\ena
where $\mu$ runs over all strict partitions.
For any permutation $\mu$ of a strict partition we define $\xi_\mu$ by Equation (\ref{eq:111}).
Then we have
\bea
&&
\xi_{(\mu_{\sigma(1)},\cdots,\mu_{\sigma(2l)})}=\sgn\sigma\,\xi_{(\mu_1,\cdots,\mu_{2l})},
\non
\ena
for a permutation $\sigma$ of degree $2l$.
In general we define $\xi_{(m_1,\cdots,m_n)}$, $m_i\ge0$, $\sharp \{ i | m_i=0\}\le 1$, such that it is skew symmetric in the indices.
It means, in particular, $\xi_{(m_1,\cdots,m_n)}=0$ if $m_i=m_j$ for some $i\neq j$.



In order to study the expansion of the addition formulae the following proposition plays a key role.
\begin{prop}
Let $\alpha_1,\cdots,\alpha_{2n}$ be parameters which satisfy $|\alpha_1|>\cdots>|\alpha_{2n}|$.
Then 
\bea
&&
A_{1\cdots 2n}\tau(x+2\sum_{i=1}^{2n}[\alpha_i]_o)=\sum_{\substack{m_i\in\mathbb{Z},i\neq 2n,\\m_{2n}\ge0}} \tilde{\xi}_{(m_1\cdots,m_{2n})}(x)\alpha_1^{m_1}\cdots\alpha_{2n}^{m_{2n}}.
\label{225}
\ena
If $m_i\ge0$ for any $i$ and the number of $i$ with $m_i=0$ is at most one, $\tilde{\xi}_{(m_1,\cdots,m_{2n})}(x)$ is skew symmetric in the indices.
If $(m_1,\cdots,m_{2n})$ is a permutation of a strict partition $\lambda$, then $\tilde{\xi}_{(m_1\cdots,m_{2n})}(0)=2^{l(\lambda)}\xi_{(m_1,\cdots,m_{2n})}$.
\end{prop}

\noindent\pf\,
We use free fermions \cite{DJKM1,DJKM2} (see appendix A for notation) to prove this proposition.
Let us consider the vertex operator
\bea
&&
X_B(\alpha)=e^{\sum_{n:odd}x_n \alpha^n} e^{-2\sum_{n:odd}\tilde{\partial}_n \alpha^{-n}}.
\non
\ena
The vertex operators satisfy, for $|\alpha|>|\beta|$,
\bea
&&
X_B(\alpha)X_B(\beta)=\frac{1-\beta/\alpha}{1+\beta/\alpha}e^{\sum x_n \alpha^n+\sum x_n \beta^n}e^{-2\sum\tilde{\partial}_n \alpha^{-n}-2\sum\tilde{\partial}_n \beta^{-n}}.
\label{eq:59}
\ena
We apply this vertex operator to $1=\langle0|e^{H_B(x)}|0\rangle$.
Using(\ref{eq:59}) we get
\bea
&&
X_B(\alpha_1)\cdots X_B(\alpha_{2n})\cdot 1=A_{1\cdots 2n}e^{\sum_{i=1}^{2n}\sum_{k:odd}x_k\alpha_i^k}.
\label{eq:60}
\ena
By the boson-fermion correspondence this is equal to 
\bea
&&
X_B(\alpha_1)\cdots X_B(\alpha_{2n})\langle0|e^{H_B(x)}|0\rangle=2^{n}\langle0|e^{H_B(x)}\phi(\alpha_1)\cdots\phi(\alpha_{2n})|0\rangle
\non
\\
&&
\hskip5.2cm=\sum_{m_i\in\mathbb{Z}} 2^{n}\langle0|e^{H_B(x)}\phi_{m_1}\cdots\phi_{m_{2n}}|0\rangle\alpha_1^{m_1}\cdots\alpha_{2n}^{m_{2n}}
\label{eq:81}
\\
&&
\hskip5.2cm=\sum_{m_i\in\mathbb{Z}} \eta_{(m_1,\cdots,m_{2n})}(x)\alpha_1^{m_1}\cdots\alpha_{2n}^{m_{2n}},
\label{eq:61}
\ena
where
\bea
&&
\eta_{(m_1,\cdots,m_{2n})}(x)=2^n\langle0|e^{H_B(x)}\phi_{m_1}\cdots\phi_{m_{2n}}|0\rangle.
\non
\ena
Therefore
\bea
&&
A_{1\cdots 2n}\tau(x+2\sum_{i=1}^{2n}[\alpha_i]_o)=A_{1\cdots 2n}e^{\sum_{i=1}^{2n}\sum_{k:odd}2\tilde{\partial}_k \alpha_i^k}\tau(x)
\non
\\
&&
\hskip4cm=\sum_{m_i\in\mathbb{Z}}\tilde{\xi}_{(m_1,\cdots,m_{2n})}(x)\alpha_1^{m_1}\cdots\alpha_{2n}^{m_{2n}},
\non
\ena
where
\bea
&&
\tilde{\xi}_{(m_1,\cdots,m_{2n})}(x)=\eta_{(m_1,\cdots,m_{2n})}(2\tilde{\partial})\tau(x).
\non
\ena
If $m_1>\cdots>m_{2n}\ge0$, it is known that (see \cite{O2, DJKM1,Y1}):
\bea
&&
\langle0|e^{H_B(x)}\phi_{m_1}\cdots\phi_{m_{2n}}|0\rangle=2^{-n}Q_{\lambda}\left(\frac{x}{2}\right),\,\,\lambda=(m_1,\cdots,m_{2n}).
\non
\ena
In this case we have, by (\ref{eq:111}),
\bea
&&
\begin{cases}
\tilde{\xi}_{(m_1,\cdots,m_{2n})}(0)=2^{2n}\xi_{(m_1,\cdots,m_{2n})},\,\,{\text{if $m_{2n}\neq0$}},
\\
\tilde{\xi}_{(m_1,\cdots,m_{2n})}(0)=2^{2n-1}\xi_{(m_1,\cdots,m_{2n})},\,\,{\text{if $m_{2n}=0$}}.
\end{cases}
\label{2000}
\ena
By the commutation relations of $\{\phi_n\}$ if there are no pairs $(i,j)$ such that $m_i+m_j=0$, then $\tilde{\xi}_{(m_1,\cdots,m_{2n})}(x)$ is skew symmetric in the indices.
In particular $\tilde{\xi}_{(m_1,\cdots,m_{2n})}(x)$ is skew symmetric if $m_i\ge0$ for any $i$ and the number of $i$ with $m_i=0$ is at most one.
Therefore similar equations to (\ref{2000}) are valid for an arbitrary permutation of $(m_1,\cdots,m_k)$, $m_1>\cdots >m_{2n}\ge0$.
Finally in the sum of (\ref{eq:61}) $m_{2n}\ge0$ since $\phi_n|0\rangle=0$ for $n<0$.\qed

\vskip5mm
\noindent{\bf{Remark.}}\,\,
Since $m_{2n}\ge0$, in the sum in the right hand side of (\ref{225}) we can put $\alpha_{2n}=0$.
Then a similar expansion is valid for $A_{1\cdots2n-1}\tau(x+\sum_{i=1}^{2n-1}[\alpha_i]_o)$ and $\tilde{\xi}_{(m_1,\cdots,m_{2n-1})}(0)=2^{2n-1}\xi_{(m_1,\cdots,m_{2n-1})}$ for $m_i\ge 1,\,1\le i\le2n-1$.

\vskip5mm

Expanding addition formulae using Proposition 1 we have
\begin{prop}
Suppose that $\tau(x)$ given by (\ref{eq:51}) is a solution of the BKP hierarchy.
Let $(n_1,\cdots,n_l)$ and $(m_1,\cdots,m_k)$ be permutations of some strict partition of length $l$ and $k$ respectively.
Then we have the following equations.

(i) For any $k,l\ge1$ such that $k+l\ge3$ is odd we have
\bea
&&
\xi_{(n_1,\cdots,n_l)}\xi_{(m_1,\cdots,m_k)}=\sum_{i=1}^l(-1)^{i-1}\xi_{(n_i,m_1,\cdots,m_k)}\xi_{(n_1,\cdots,\hat{n}_i,\cdots,n_l)}
\non
\\
&&
\hskip3.5cm+\sum_{i=1}^k(-1)^{l+i-1}\xi_{(m_1,\cdots,\hat{m}_i,\cdots,m_k)}\xi_{(n_1,\cdots,n_l,m_i)}.
\label{223}
\ena

(ii) For any $k,l\ge1$ such that $k+l\ge 4$ is even we have
\bea
&&
\xi_{(n_1,\cdots,n_{l})}\xi_{(m_1,\cdots,m_k)}=\sum_{i=1}^{l-1}(-1)^{k+i-1}\xi_{(n_i,n_{l},m_1,\cdots,m_k)}\xi_{(n_1,\cdots,\hat{n}_i,\cdots,n_{l-1})}
\non
\\
&&
\hskip3.5cm +\sum_{i=1}^k(-1)^{i-1}\xi_{(n_{l},m_1,\cdots,\hat{m}_i,\cdots,m_k)}\xi_{(n_1,\cdots,n_{l-1},m_i)}.
\label{224}
\ena
\end{prop}

\noindent\pf\,
We prove (ii).
Equation (i) can be proved in a similar way.

Assume that  $\tau(x)$ is a solution of the BKP hierarchy.
We set, in (\ref{212}), $n=k+l$ and 
\bea
&&
\alpha_{l}=-\beta_1,\cdots,\alpha_{n-1}=-\beta_k.
\non
\ena
In the following we write $\alpha_n$ by $\alpha_l$.
Then shift $x$ to $x+2\sum_{j=1}^k[\beta_j]_o$.
Multiplying the resulting equation by 
\bea
&&
\prod_{s=1}^{l}\prod_{r=1}^k(-1)^k\frac{\alpha_s-\beta_r}{\alpha_s+\beta_r},
\non
\ena
we have
\bea
&&
A_{1\cdots l}B_{1\cdots k}\tau(x+2\sum_{j=1}^k[\beta_j]_o)\tau(x+2\sum_{j=1}^l[\alpha_j]_o)
\non
\\
&&
=\sum_{i=1}^{l-1}(-1)^{i-1}A_{1\cdots\hat{i}\cdots l-1}B_{1\cdots k}\frac{\alpha_{i,l}}{\tilde{\alpha}_{i,l}}\prod_{r=1}^k\left(\frac{\alpha_i-\beta_r}{\alpha_i+\beta_r}\cdot\frac{\beta_r-\alpha_l}{\beta_r+\alpha_l}\right)
\non
\\
&&
\times\tau(x+2[\alpha_i]_o+2\sum_{j=1}^k[\beta_j]_o+2[\alpha_l]_o)\tau(x+2\sum_{j\neq i}^{l-1}[\alpha_j]_o)
\non
\\
&&
+\sum_{i=1}^k(-1)^{l+i}A_{1\cdots l-1}B_{1\cdots\hat{i}\cdots k}\prod_{s=1}^{l-1}\frac{\alpha_s-\beta_i}{\alpha_s+\beta_i}\prod_{r=1,r\neq i}^{k}\frac{\beta_r-\alpha_l}{\beta_r+\alpha_l}
\non
\\
&&
\times\tau(x+2\sum_{j\neq i}^k[\beta_j]_o+2[\alpha_l]_o)\tau(x+2\sum_{j=1}^{l-1}[\alpha_j]_o+2[\beta_i]_o),
\label{226}
\ena
where 
\bea
&&
B_{1\cdots k}=\prod_{i<j}^k\frac{\beta_i-\beta_j}{\beta_i+\beta_j}.
\non
\ena
We assume $|\alpha_1|>\cdots>|\alpha_{l-1}|>|\beta_1|>\cdots>|\beta_k|>|\alpha_l|$.
Set $x=0$ and expand (\ref{226}) using (\ref{225}), then we get
\bea
&&
\sum_{n_s,m_r\in\mathbb{Z}}\tilde{\xi}_{(n_1,\cdots,n_l)}(0)\tilde{\xi}_{(m_1,\cdots,m_k)}(0)\alpha_1^{n_1}\cdots\alpha_l^{n_l}\beta_1^{m_1}\cdots\beta_k^{m_k}
\non
\\
&&
=\sum_{i=1}^{l-1}(-1)^{k+i-1}\sum_{n_s,m_r\in\mathbb{Z}}\tilde{\xi}_{(n_1,\cdots,\hat{n}_i,\cdots,n_{l-1})}(0)\tilde{\xi}_{(n_i,n_l,m_1,\cdots,m_k)}(0)\alpha_1^{n_1}\cdots\alpha_l^{n_l}\beta_1^{m_1}\cdots\beta_k^{m_k}
\non
\\
&&
+\sum_{i=1}^k(-1)^{i-1}\sum_{n_s,m_r\in\mathbb{Z}}\tilde{\xi}_{(n_1,\cdots,n_{l-1},m_i)}(0)\tilde{\xi}_{(n_l,m_1,\cdots,\hat{m}_i,\cdots,m_k)}(0)\alpha_1^{n_1}\cdots\alpha_l^{n_l}\beta_1^{m_1}\cdots\beta_k^{m_k}.
\non
\ena
Compare the coefficients of $\alpha_1^{n_1}\cdots\alpha_l^{n_l}\beta_1^{m_1}\cdots\beta_k^{m_k}$, $m_i,n_j\ge0$ and rewrite them in terms of $\xi$ using (\ref{2000}).
Then some powers of 2 appear in both sides, however they are canceled. 
Thus we obtain (\ref{224}).
\qed

\vskip5mm
The following proposition had been proved in \cite{DJKM2}.
\begin{prop}{\bf\cite{DJKM2}}
A series $\tau(x)$ of the form (\ref{eq:51}) is a solution of the BKP hierarchy if and only if the coefficients $\{\xi_{\lambda}\}$ satisfy the following equations:
\bea
&&
\xi_{(m_1,\cdots,m_k)}\xi_{(m_1,\cdots,m_k,n_1,n_2,n_3,n_4)}=\xi_{(m_1,\cdots,m_k,n_1,n_4)}\xi_{(m_1,\cdots,m_k,n_2,n_3)}
\non
\\
&&
\hskip5.5cm-\xi_{(m_1,\cdots,m_k,n_2,n_4)}\xi_{(m_1,\cdots,m_k,n_1,n_3)}
\non
\\
&&
\hskip5.5cm+\xi_{(m_1,\cdots,m_k,n_3,n_4)}\xi_{(m_1,\cdots,m_k,n_1,n_2)},
\label{3210}
\ena
where $k$ is even and $(m_1,\cdots,m_k)$ and $(m_1,\cdots,m_k,n_1,n_2,n_3,n_4)$ are permutations of some strict partitions $\lambda$ and $\mu$ respectively.
\end{prop}

\begin{cor}
Suppose that the coefficients $\{\xi_{\mu}\}$ satisfy (\ref{223}) and (\ref{224}), then $\tau(x)$ is a solution of the BKP hierarchy.
\end{cor}

\noindent\pf\,
If $m_i\neq0,\,n_j\neq0$ for any $i,j$, set $l=k+4$ and $n_1=m_1,\cdots,n_k=m_k$ in (\ref{224}).
Then the second summation of (\ref{224}) becomes zero and three terms of the first summation are left.
Thus we obtain (\ref{3210}).
Other cases are similarly proved using (\ref{224}) or (\ref{223}).\qed

Therefore the set of equations (\ref{223}) and (\ref{224}) are also equivalent to the BKP hierarchy.

\section{Expansion coefficients of $\tau(x)$}
In this section we study the expansion coefficients of $\tau(x)$ in detail.

We first introduce notation and some properties on Pfaffians.
Let A$=(a_{ij})_{1\le i,j\le 2m}$ be a skew-symmetric matrix.
Then the Pfaffian $\Pf(a_{ij})$ \cite{H1} is defined by
\bea
&&
\Pf(a_{ij})=\sum \sgn(i_1,\dots,i_{2m})\cdot a_{i_1,i_2}a_{i_3,i_4}\cdots a_{i_{2m-1},i_{2m}},
\label{**}
\ena
where the sum is over all permutations of (1,\dots,2m) such that
\bea
&&
i_1<i_3<\cdots<i_{2m-1},\hskip3mm i_1<i_2,\cdots,i_{2m-1}<i_{2m},
\non
\ena
and $\sgn(i_1,\dots,i_{2m})$ is the signature of the permutation $(i_1,\dots,i_{2m})$.
In order to describe $\Pf(a_{ij})$ more conveniently we use some set of symbols $X_i,\,1\le i\le 2m$ (see Theorem 3).
First we set $(X_i,X_j)=a_{ij}$.
For any permutation $i_1,\cdots,i_{2m}$ of $1,\cdots,2m$ we define
\bea
&&
(X_{i_1},\cdots,X_{i_{2m}})=\Pf ((X_{i_k},X_{i_l}))_{1\le k,l\le 2m}.
\non
\ena
Then it is skew symmetric in the indices.
The Pfaffian has the following expansion:
\bea
&&
(X_1,\cdots,X_{2m})=\sum_{j=2}^{2m}(-1)^j(X_1,X_j)(X_2,\dots,\hat{X_j},\dots,X_{2m}).
\non
\ena
For example, in the case of $m=2$, 
\bea
&&
(X_1,X_2,X_3,X_4)=(X_1,X_2)(X_3,X_4)-(X_1,X_3)(X_2,X_4)+(X_1,X_4)(X_2,X_3).
\non
\ena

A Pfaffian analogue of Pl\"{u}cker relations is known \cite{O1}:
\bea
&&
\sum_{r=1}^R(-1)^r(X_{i_1},\cdots,X_{i_S},X_{j_r})(X_{j_1},\cdots,\hat{X}_{j_r},\cdots,X_{j_R})
\non
\\
&&
+\sum_{s=1}^S(-1)^s(X_{i_1}\cdots,\hat{X}_{i_s},\cdots,X_{i_S})(X_{j_1},\cdots,X_{j_R},X_{i_s})=0,
\label{233}
\ena
where $R$ and $S$ are odd.

In \cite{DJKM2} the following theorem is proved by solving Equations (\ref{3210}).
\begin{theorem}{\bf\cite{DJKM2}}
A formal power series $\tau(x)$ given by (\ref{eq:51}) satisfying the condition $\tau(0)=1$ is a solution of the BKP hierarchy if and only if the coefficients $\{\xi_{\mu}\}$ satisfy
\bea
&&
\xi_{\mu}=\Pf\left(\xi_{(\mu_i,\mu_j)}\right)_{1\le i,j \le 2n}.
\label{2300}
\ena
\end{theorem}

We study the case of $\tau(0)=0$.
In this case it seems that it is difficult to derive the formula corresponding to (\ref{2300}) by using Equations (\ref{3210}) only.
Our strategy is to use larger set of Equations (\ref{223}) and (\ref{224}).
Let $\lambda=(\lambda_1,\cdots,\lambda_M)$ be a strict partition of length $M$.
We assume that $\tau(x)$ has the following expansion:
\bea
&&
\tau(x)=Q_{\lambda}(\frac{x}{2})+\sum_{|\mu|>|\lambda|}\xi_{\mu}Q_{\mu}(\frac{x}{2}).
\label{230}
\ena
We set $\xi_\mu=0$ if $|\mu|\le|\lambda|$ and $\mu\neq\lambda$.
We consider the following subset of the non-trivial expansion coefficients in (\ref{230}),
\bea
&&
\begin{cases}
\xi_{(\lambda_1,\cdots,\hat{\lambda}_i,\cdots,\lambda_M,n)},\,\,\,n>\lambda_i,\,n\neq\lambda_j\,{\text{for any $j$}},
\\
\xi_{(\lambda_1,\cdots,\lambda_M,n)},\,\,\,n\ge 1,n\neq\lambda_i\,\,{\text{for any $i$}},
\\
\xi_{(\lambda_1,\cdots,\lambda_M,n_i,n_j)},\,\,\,n_i>n_j\ge 1,n_i,n_j\neq\lambda_i\,\,{\text{for any $i$}}.
\end{cases}
\label{*}
\ena

We define the components of Pfaffian as
\bea
&&
(\Lambda^{(i)},n)=\xi_{(\lambda_1,\cdots,\hat{\lambda}_i,\cdots,\lambda_M,n)},
\non
\\
&&
(\Lambda,n)=\xi_{(\lambda_1,\cdots,\lambda_M,n)},\,\,\,n\ge 1,
\non
\\
&&
(n_i,n_j)=\xi_{(\lambda_1,\cdots,\lambda_M,n_i,n_j)},\,\,\,n_i>n_j\ge 1,
\non
\\
&&
(\Lambda,\Lambda^{(i)})=(\Lambda^{(i)},\Lambda^{(j)})=0.
\non
\ena

\begin{ex}
(i)\,Pfaffian $(\Lambda,\Lambda^{(1)},n_1,n_2)$ is expanded as
\bea
(\Lambda,\Lambda^{(1)},n_1,n_2)&=&(\Lambda,\Lambda^{(1)})(n_1,n_2)-(\Lambda,n_1)(\Lambda^{(1)},n_2)+(\Lambda,n_2)(\Lambda^{(1)},n_1)
\non
\\
&=&-\xi_{(\lambda_1,\cdots,\lambda_M,n_1)}\xi_{(\lambda_2,\cdots,\lambda_M,n_2)}+\xi_{(\lambda_1,\cdots,\lambda_M,n_2)}\xi_{(\lambda_2,\cdots,\lambda_M,n_1)}.
\non
\ena
(ii)\,Pfaffian $(\Lambda^{(1)},n_1,n_2,n_3)$ is expanded as
\bea
(\Lambda^{(1)},n_1,n_2,n_3)&=&(\Lambda^{(1)},n_1)(n_2,n_3)-(\Lambda^{(1)},n_2)(n_1,n_3)+(\Lambda^{(1)},n_3)(n_1,n_2)
\non
\\
&=&\xi_{(\lambda_2,\cdots,\lambda_M,n_1)}\xi_{(\lambda_1,\cdots,\lambda_M,n_2,n_3)}-\xi_{(\lambda_2,\cdots,\lambda_M,n_2)}\xi_{(\lambda_1,\cdots,\lambda_M,n_1,n_3)}
\non
\\
& &+\xi_{(\lambda_2,\cdots,\lambda_M,n_3)}\xi_{(\lambda_1,\cdots,\lambda_M,n_1,n_2)}.
\non
\ena
\end{ex}

Notice that $(\Lambda^{(i)},n)=0$ for $n<\lambda_i$, since $\sum_{j\neq i}\lambda_j+n<|\lambda|$. 

Then our main theorem is
\begin{theorem}
Suppose that $\tau(x)$ has the expansion (\ref{230}).
Then $\tau(x)$ is a solution of the BKP hierarchy if and only if the coefficients $\xi_{\mu},\,\mu=(\mu_1,\cdots,\mu_k),\,l(\mu)=k$ are given by the following formulae where the quantities in (\ref{*}) are arbitrary.

(i) $M=2L-1$,
\bea
&&
\xi_{\mu}=
\begin{cases}
(\Lambda^{(1)},\cdots,\Lambda^{(2L-1)},\mu_1,\cdots,\mu_{2l-1}),\,\,{\text{if $k=2l-1$}},\\
(\Lambda,\Lambda^{(1)},\cdots,\Lambda^{(2L-1)},\mu_1,\cdots,\mu_{2l}),\,\,{\text{if $k=2l$}}.
\end{cases}
\label{231}
\ena

(ii) $M=2L$,
\bea
&&
\xi_{\mu}=
\begin{cases}
(\Lambda,\Lambda^{(1)},\cdots,\Lambda^{(2L)},\mu_1,\cdots,\mu_{2l-1}),\,\,{\text{if $k=2l-1$}},\\
(\Lambda^{(1)},\cdots,\Lambda^{(2L)},\mu_1,\cdots,\mu_{2l}),\,\,{\text{if $k=2l$}}.
\end{cases}
\label{232}
\ena
\end{theorem}

\vskip5mm
We first remark that (\ref{231}) and (\ref{232}) are trivial equations for quantities in (\ref{*}).

Whether the quantities $\{\xi_\mu\}$ given by (\ref{231}) and (\ref{232}) satisfy $\xi_\mu=0$ for $|\mu|\le|\lambda|,\,\mu\neq\lambda$ is not very obvious.
So let us prove the following lemma.

\begin{lemma}
The quantities $\{\xi_{\mu}\}$ given by (\ref{231}) and (\ref{232}) satisfy $\xi_{\mu}=0$ if $|\mu|\le|\lambda|$ and $\mu\neq\lambda$.
\end{lemma}

\noindent\pf\,
We prove the lemma in the case of the length of $\lambda$ and $\mu$ are odd, that is the first case of (\ref{231}).
Set $\lambda=(\lambda_1,\cdots,\lambda_{2L-1})$ and $\mu=(\mu_1,\cdots,\mu_{2l-1})$.
If $l<L$, then 
\bea
&&
\xi_{(\mu_1,\cdots,\mu_{2l-1})}=(\Lambda^{(1)},\cdots,\Lambda^{(2L-1)},\mu_1,\cdots,\mu_{2l-1})
\non
\\
&&
\hskip2cm=\sum sgn(i_1,\cdots,i_{2l-1})(\Lambda^{(1)},\mu_{i_1})\cdots(\Lambda^{(2l-1)},\mu_{i_{2l-1}})(\Lambda^{(2l)},\cdots,\Lambda^{(2L-1)}).
\non
\ena
Since $(\Lambda^{(2l)},\cdots,\Lambda^{(2L-1)})=0$, $\xi_{(\mu_1,\cdots,\mu_{2l-1})}$ is zero.

If $l\ge L$, then 
\bea
&&
\xi_{(\mu_1,\cdots,\mu_{2L-1})}
\non
\\
&&
=\sum sgn(i_1,\cdots,i_{2L-1})(\Lambda^{(1)},\mu_{i_1})\cdots(\Lambda^{(2L-1)},\mu_{i_{2L-1}})(\mu_{i_{2L}},\cdots,\mu_{i_{2l-1}}).
\label{2311}
\ena
If $\lambda_j>\mu_{i_j}$ for some $j$, then $(\Lambda^{(j)},\mu_{i_j})=0$.
So if the term in (\ref{2311}) corresponding to $(i_1,\cdots,i_{2L-1})$ is not zero, then $\mu_{i_j}\ge\lambda_j$ for any $1\le j\le 2L-1$.
Since $|\mu|\le|\lambda|$, this is possible if and only if $l=L$ and $\mu_{i_j}=\lambda_j$ for any $j$.
Then $(i_1,\cdots,i_{2L-1})=(1,\cdots,2L-1)$ and $\lambda\neq\mu$ because $\lambda$ and $\mu$ are strict partitions.
So if $\lambda\neq\mu$, then $\xi_\mu=0$.
We can prove the other cases in a similar way.\qed

\vskip5mm
\noindent{\it{Proof of Theorem 3.}}\,
We first prove that if $\{\xi_{\mu}\}$ is given by (\ref{231}) and (\ref{232}), then $\tau(x)$ given by (\ref{230}) satisfies (\ref{223}) and (\ref{224}).

Consider the case of (\ref{231}).
In (\ref{233}) take $S=2L+2k'-1,\,k'\ge 1$, $R=2L+2l'-1,\,l'\ge 1$ and set 
\bea
&&
X_{i_1}=\Lambda,X_{i_2}=\Lambda^{(1)},\cdots,X_{i_{2L}}=\Lambda^{(2L-1)},X_{i_{2L+1}}=m_1,\cdots,X_{i_{2L+2k'-1}}=m_{2k'-1},
\non
\\
&&
X_{j_1}=\Lambda^{(1)},\cdots,X_{j_{2L-1}}=\Lambda^{(2L-1)},X_{j_{2L}}=n_1,\cdots,X_{j_{2L+2l'-1}}=n_{2l'},
\non
\ena
then we have (\ref{223}) with $k=2k'-1$ and $l=2l'$.
We can get (\ref{223}) with $k=2k'$ and $l=2l'-1$ by setting 
\bea
&&
X_{i_1}=\Lambda^{(1)},\cdots,X_{i_{2L-1}}=\Lambda^{(2L-1)},X_{i_{2L}}=m_1,\cdots,X_{i_{2L+2k'-1}}=m_{2k'},
\non
\\
&&
X_{j_1}=\Lambda,X_{i_2}=\Lambda^{(1)},\cdots,X_{j_{2L}}=\Lambda^{(2L-1)},X_{j_{2L+1}}=n_1,\cdots,X_{j_{2L+2l'-1}}=n_{2l'-1}.
\non
\ena
In (\ref{233}) take $S=2L+2k'+1,k'\ge1$, $R=2L+2l'-1,l'\ge1$ and set 
\bea
&&
X_{i_1}=\Lambda,X_{i_2}=\Lambda^{(1)},\cdots,X_{i_{2L}}=\Lambda^{(2L-1)},
\non
\\
&&
X_{i_{2L+1}}=m_1,\cdots,X_{i_{2L+2k'}}=m_{2k'},X_{i_{2L+2k'+1}}=n_{2l'},
\non
\\
&&
X_{j_1}=\Lambda,X_{j_2}=\Lambda^{(1)},\cdots,X_{j_{2L}}=\Lambda^{(2L-1)},X_{j_{2L-1}}=n_1,\cdots,X_{j_{2L+2l'-1}}=n_{2l'-1},
\non
\ena
then we have (\ref{224}) with $k=2k'$ and $l=2l'$.
We can get (\ref{224}) with $k=2k'-1$ and $l=2l'-1$ by setting, in (\ref{233}), 
\bea
&&
X_{i_1}=\Lambda^{(1)},\cdots,X_{i_{2L-1}}=\Lambda^{(2L-1)},
\non
\\
&&
X_{i_{2L}}=m_1,\cdots,X_{i_{2L+2k'-2}}=m_{2k'-1},X_{i_{2L+2k'-1}}=n_{2l'-1},
\non
\\
&&
X_{j_1}=\Lambda^{(1)},\cdots,X_{j_{2L-1}}=\Lambda^{(2L-1)},X_{j_{2L}}=n_1,\cdots,X_{j_{2L+2l'-3}}=n_{2l'-2}.
\non
\ena

Next consider $\xi_\mu$ given by (\ref{232}).
In (\ref{233}) with $S=2L+2k'-1,\,k'\ge 1$, $R=2L+2l'+1,\,l'\ge1$ set
\bea
&&
X_{i_1}=\Lambda^{(1)},\cdots,X_{i_{2L}}=\Lambda^{(2L)},X_{i_{2L+1}}=m_1,\cdots,X_{i_{2L+2k'-1}}=m_{2k'-1},
\non
\\
&&
X_{j_1}=\Lambda,X_{j_2}=\Lambda^{(1)},\cdots,X_{j_{2L+1}}=\Lambda^{(2L)},X_{j_{2L+2}}=n_1,\cdots,X_{j_{2L+2l'+1}}=n_{2l'},
\non
\ena
then we have (\ref{223}) with $k=2k'-1$ and $l=2l'$.
We have (\ref{223}) with $k=2k'$ and $l=2l'-1$ by setting 
\bea
&&
X_{i_1}=\Lambda,X_{i_2}=\Lambda^{(1)},\cdots,X_{i_{2L+1}}=\Lambda^{(2L)},X_{i_{2L+2}}=m_1,\cdots,X_{i_{2L+2k'+1}}=m_{2k'},
\non
\\
&&
X_{j_1}=\Lambda^{(1)},\cdots,X_{j_{2L}}=\Lambda^{(2L)},X_{j_{2L+1}}=n_1,\cdots,X_{j_{2L+2l'-1}}=n_{2l'-1}.
\non
\ena
We set, in (\ref{233}) , $S=2L+2k'+1,k'\ge 1,\,R=2L+2l'-1,l'\ge 1$, and 
\bea
&&
X_{i_1}=\Lambda^{(1)},\cdots,X_{i_{2L}}=\Lambda^{(2L)},
\non
\\
&&
X_{i_{2L+1}}=m_1,\cdots,X_{i_{2L+2k'}}=m_{2k'},X_{i_{2L+2k'+1}}=n_{2l'},
\non
\\
&&
X_{j_1}=\Lambda^{(1)},\cdots,X_{j_{2L}}=\Lambda^{(2L)},X_{j_{2L+1}}=n_1,\cdots,X_{j_{2L+2l'-1}}=n_{2l'-1},
\non
\ena
then we have (\ref{224}) with $k=2k'$ and $l=2l'$.
Similarly we have (\ref{224}) with $k=2k'-1$ and $l=2l'-1$ by setting 
\bea
&&
X_{i_1}=\Lambda,X_{i_2}=\Lambda^{(1)},\cdots,X_{i_{2L+1}}=\Lambda^{(2L)},
\non
\\
&&
X_{i_{2L+2}}=m_1,\cdots,X_{i_{2L+2k'}}=m_{2k'-1},X_{i_{2L+2k'+1}}=n_{2l'-1},
\non
\\
&&
X_{j_1}=\Lambda,X_{j_2}=\Lambda^{(1)},\cdots,X_{j_{2L+1}}=\Lambda^{(2L)},X_{j_{2L+2}}=n_1,\cdots,X_{j_{2L+2l'-1}}=n_{2l'-2}.
\non
\ena
By Corollary 1 $\tau(x)$ is a solution of the BKP hierarchy.

\vskip5mm
Conversely we prove that if $\tau(x)$ is a solution of the BKP hierarchy, then $\{\xi_{\mu}\}$ is given by (\ref{231}) and (\ref{232}).
Let us prove (i).
The proof of (ii) is similar. 
We use the following special cases of Equations (\ref{223}) and (\ref{224}).

In (\ref{223}) with $l$ being even and in (\ref{224}) with $l$ being odd we take $k=2L-1$ and $(m_1,\cdots,m_k)=(\lambda_1,\cdots,\lambda_{2L-1})$.
Then we have the following relations.
If $l$ is even, then we have
\bea
&&
\xi_{(n_1,\cdots,n_l)}=\sum_{i=1}^l(-1)^{i-1}\xi_{(n_i,\lambda_1,\cdots,\lambda_{2L-1})}\xi_{(n_1,\cdots,\hat{n}_i,\cdots,n_l)}.
\label{234}
\ena
If $l$ is odd, we have 
\bea
&&
\xi_{(n_1,\cdots,n_{l})}=\sum_{i=1}^{l-1}(-1)^{i}\xi_{(n_i,n_{l},\lambda_1,\cdots,\lambda_{2L-1})}\xi_{(n_1,\cdots,\hat{n}_i,\cdots,n_{l-1})}
\non
\\
&&
\hskip2cm+\sum_{i=1}^{2L-1}(-1)^{i-1}\xi_{(n_{l},\lambda_1,\cdots,\hat{\lambda}_i,\cdots,\lambda_{2L-1})}\xi_{(n_1,\cdots,n_{l-1},\lambda_i)}.
\label{2301}
\ena

We first prove the following lemmas.

\begin{lemma}
For $k\ge2$ and $n\ge 1$, we have
\bea
&&
\xi_{(\lambda_1,\cdots,\hat{\lambda}_{i_1},\cdots,\hat{\lambda}_{i_k},\cdots,\lambda_{2L-1},n)}=0,
\label{235}
\ena
where $\hat{\lambda}_{i_j}$ means to remove $\lambda_{i_j}$.
\end{lemma}

\noindent\pf\,
We assume $L\ge2$.
Set $(\tilde{\lambda}_1,\cdots,\tilde{\lambda}_{2L-k-1},n)=(\lambda_1,\cdots,\hat{\lambda}_{i_1},\cdots,\hat{\lambda}_{i_k},\cdots,\lambda_{2L-1},n)$.
Let us consider the case of $k=2L-1$, namely $(\tilde{\lambda}_1,\cdots,\tilde{\lambda}_{2L-k-1})=\phi$.
Set $l=1$ in (\ref{2301}), then we have
\bea
&&
\xi_{(n)}=\sum_{i=1}^{2L-1}(-1)^{i-1}\xi_{(n,\lambda_1,\cdots,\hat{\lambda}_i,\cdots,\lambda_{2L-1})}\xi_{(\lambda_i)}.
\label{2302}
\ena
Here if $|\mu|<|\lambda|$, $\xi_{\mu}=0$ by the assumption.
Then we get $\xi_{(n)}=0$ since $\xi_{(\lambda_i)}=0$ if $L\ge 2$.
Thus (\ref{235}) is valid for the case $k=2L-1$.

We prove (\ref{235}) if $k$ is even.
By (\ref{234}) we have
\bea
&&
\xi_{(\tilde{\lambda}_1,\cdots,\tilde{\lambda}_{2L-k-1},n)}=\sum_{i=1}^{2L-k-1}(-1)^{i-1}\xi_{(\tilde{\lambda}_i,\lambda_1,\cdots,\lambda_{2L-1})}\xi_{(\tilde{\lambda}_1,\cdots,\hat{\tilde{\lambda}}_i,\cdots,\tilde{\lambda}_{2L-k-1},n)}
\non
\\
&&
\hskip3cm+(-1)^{2L-k-1}\xi_{(n,\lambda_1,\cdots,\lambda_{2L-1})}\xi_{(\tilde{\lambda}_1,\cdots,\tilde{\lambda}_{2L-k-1})}.
\label{236}
\ena
Here $\xi_{(\tilde{\lambda}_i,\lambda_1,\cdots,\lambda_{2L-1})}=0$ for any $i$, since $\lambda_i$ appears twice.
We also have $\xi_{(\tilde{\lambda}_1,\cdots,\tilde{\lambda}_{2L-k-1})}=0$, since $\sum_{i=1}^{2L-k-1}\tilde{\lambda}_i<|\lambda|$.
Thus $\xi_{(\tilde{\lambda}_1,\cdots,\hat{\tilde{\lambda}}_i,\cdots,\tilde{\lambda}_{2L-k-1},n)}=0$.
If $k$ is odd and $k<2L-1$, it is possible to prove (\ref{235}) in a similar manner.\qed

\begin{lemma}
The quantity $\xi_{(n_1,\cdots,n_l)}$ is uniquely determined if elements in (\ref{*}) are taken as an initial condition.
Moreover $\xi_{(n_1,\cdots,n_l)}$ is expressed as a polynomial in elements in (\ref{*}).
\end{lemma}

\noindent\pf\,
We prove this statement by induction on $l$.
We first prove the case of $l=2$.
Set $l=2$ in (\ref{234}).
Then we have
\bea
&&
\xi_{(n_1,n_2)}=\xi_{(n_1,\lambda_1,\cdots,\lambda_{2L-1})}\xi_{(n_2)}- 
\xi_{(n_2,\lambda_1,\cdots,\lambda_{2L-1})}\xi_{(n_1)}.
\label{237}
\ena
In case of $L=1$, namely $\lambda=(\lambda_1)$, (\ref{237}) becomes
\bea
&&
\xi_{(n_1,n_2)}=\xi_{(n_1,\lambda_1)}\xi_{(n_2)}- 
\xi_{(n_2,\lambda_1)}\xi_{(n_1)}.
\non
\ena
In the right hand side $\xi_{(\lambda_1,n_i)}=-\xi_{(n_i,\lambda_1)}$ and $\xi_{(n_i)},\,n_i>\lambda_1$ are contained in (\ref{*}) and $\xi_{(n_i)}=0$ for $n_i<\lambda_1$, $\xi_{(n_i)}=1$ for $n_i=\lambda_1$,
Thus the right hand side of (\ref{237}) is expressed as a polynomial of elements in (\ref{*}).
Thus the case of $l=2$ of the lemma is proved. 

Let us consider the case $l\ge2$.
Suppose that Proposition 2 is valid for any $l'\le l$.
If $l$ is odd, we have, by (\ref{234}),
\bea
&&
\xi_{(n_1,\cdots,n_{l+1})}=\sum_{i=1}^{l+1}(-1)^{i-1}\xi_{(n_i,\lambda_1,\cdots,\lambda_{2L-1})}\xi_{(n_1,\cdots,\hat{n}_i,\cdots,n_{l+1})}
\non
\\
&&
\hskip2cm=\sum_{i=1}^{l+1}(-1)^{i}(\Lambda,n_i)\xi_{(n_1,\cdots,\hat{n}_i,\cdots,n_{l+1})}.
\non
\ena
In the right hand side $(\Lambda,n_i)$ is contained in (\ref{*}) and $\xi_{(n_1,\cdots,\hat{n}_i,\cdots,n_{l+1})}$ is a polynomial of elements in (\ref{*}) by induction hypothesis.
Thus $\xi_{(n_1,\cdots,n_{l+1})}$ is expressed as a polynomial of elements in (\ref{*}) in this case. 

If $l$ is even, using (\ref{2301}) we have
\bea
&&
\xi_{(n_1,\cdots,n_{l+1})}=\sum_{i=1}^l (-1)^{i}\xi_{(n_i,n_{l+1},\lambda_1,\cdots,\lambda_{2L-1})}\xi_{(n_1,\cdots,\hat{n}_i,\cdots,n_{l})}
\non
\\&&
\hskip2cm+\sum_{i=1}^{2L-1}(-1)^{i-1}\xi_{(n_{l+1},\lambda_1,\cdots,\hat{\lambda}_i,\cdots,\lambda_{2L-1})}\xi_{(n_1,\cdots,n_{l},\lambda_i)}
\non
\\
&&
\hskip2cm=\sum_{i=1}^l (-1)^{i}(n_i,n_{l+1})\xi_{(n_1,\cdots,\hat{n}_i,\cdots,n_{l})}
\non
\\
&&
\hskip2cm+\sum_{i=1}^{2L-1}(-1)^{i-1}(\Lambda^{(i)},n_{l+1})\xi_{(n_1,\cdots,n_{l},\lambda_i)}.
\label{238}
\ena
In the right hand side $(n_i,n_{l+1})$ and $(\Lambda^{(i)},n_{l+1})$ are in (\ref{*}) or 0 and $\xi_{(n_1,\cdots,\hat{n}_i,\cdots,n_{l})}$ is a polynomial of elements in (\ref{*}) by the assumption of induction.
We shall prove that $\xi_{(n_1,\cdots,n_{l},\lambda_i)}$ can be expressed as a polynomial of elements from (\ref{*}).

Setting $n_1=\lambda_{i_1},\,i_1\in\{1,\cdots,2L-1\}$ in (\ref{238}) we have
\bea
&&
\xi_{(\lambda_{i_1},n_2,\cdots,n_{l+1})}=\sum_{i=2}^l (-1)^{i}\xi_{(n_i,n_{l+1},\lambda_1,\cdots,\lambda_{2L-1})}\xi_{(\lambda_{i_1},n_2,\cdots,\hat{n}_i,\cdots,n_{l})}
\non
\\
&&
\hskip2.5cm+\sum_{i=1}^{2L-1}(-1)^{i-1}\xi_{(n_{l+1},\lambda_1,\cdots,\hat{\lambda}_i,\cdots,\lambda_{2L-1})}\xi_{(\lambda_{i_1},n_2,\cdots,n_{l},\lambda_i)}.
\label{240}
\ena
Then it is sufficient to prove that $\xi_{(\lambda_{i_1},n_2,\cdots,n_l,\lambda_j)}$ can be expressed as a polynomial of elements in (\ref{*}).

Setting $n_2=\lambda_{i_2},\,i_2\neq i_1,\,i_2\in\{1,\cdots,2L-1\}$ in (\ref{240}), we have
\bea
&&
\xi_{(\lambda_{i_1},\lambda_{i_2},\cdots,n_{l+1})}=\sum_{i=3}^l (-1)^{i}\xi_{(n_i,n_{l+1},\lambda_1,\cdots,\lambda_{2L-1})}\xi_{(\lambda_{i_1},\lambda_{i_2},n_3,\cdots,\hat{n}_i,\cdots,n_{l})}
\non
\\
&&
\hskip2.5cm+\sum_{i=1}^{2L-1}(-1)^{i-1}\xi_{(n_{l+1},\lambda_1,\cdots,\hat{\lambda}_i,\cdots,\lambda_{2L-1})}\xi_{(\lambda_{i_1},\lambda_{i_2},n_3,\cdots,n_{l},\lambda_i)}.
\label{241}
\ena
Repeating this, if $l+1>2L-1$, we finally arrive at the quantity of the form $\xi_{(\lambda_{i_1},\cdots,\lambda_{i_{l+1}})}$ which is zero, since some $\lambda_{i_j}$ appears more than once.
If $l+1=2L-1$, then we come to 0 or $\xi_{(\lambda_1,\cdots,\lambda_{2L-1})}=\pm1$ where $(i_1,\cdots,i_{2L-1})$ is a permutation of $(1,\cdots,2L-1)$.
If $l+1<2L-1$, then we have $\xi_{(\lambda_{i_1},\cdots,\lambda_{i_{l+1}})}=0$ since the weight of $(\lambda_{i_1},\cdots,\lambda_{i_{l+1}})$ is less than $|\lambda'|$.
Therefore $\xi_{(n_1,\cdots,n_{l+1})}$ can be expressed as a polynomial of elements from (\ref{*}).
The argument above shows that $\xi_{(n_1,\cdots,n_l)}$ is uniquely determined from elements in (\ref{*}).
Thus Lemma 2 is proved.\qed

\vskip5mm 
Let us return to the proof of Theorem 3.
By Lemma 3 $\xi_{(n_1,\cdots,n_l)}$ is uniquely solvable from elements in (\ref{*}) by using (\ref{234}) and (\ref{2301}).
Since $\{\xi_{\mu}\}$ given by (\ref{231}) are determined from elements in (\ref{*}) and satisfy (\ref{234}) and (\ref{2301}), $\xi_{(n_1,\cdots,n_l)}$ should coincide with that given by (\ref{231}).
The proof in case of $\lambda=(\lambda_1,\cdots,\lambda_{2L})$ is similar.\qed

\vskip5mm
\noindent{\bf{Remark.}}\,\,
(1) It is conjectured that $\xi_{\mu}=0$ if $\mu$ does not contain $\lambda$.
We have checked this property in the case of $\lambda=(\lambda_1)$ and $\lambda=(\lambda_1,\lambda_2)$.
We can for the moment prove the following Proposition 4 which shows a special case of this property.\\
(2) In the proof of Lemma 3 we only use the condition that $\xi_\mu=0$ if $|\mu|<|\lambda|$ and $\xi_\lambda=1$.
This fact and Lemma 1 imply that if a solution of the BKP hierarchy satisfies 
\bea
&&
\tau(x)=\sum_{|\mu|=|\lambda|}\xi_\mu Q_\mu(\frac{x}{2})+\sum_{|\mu|>|\lambda|}\xi_\mu Q_\mu(\frac{x}{2}),
\non
\ena
and $\xi_\lambda=1$, then $\xi_\mu=0$ for any $\mu\neq\lambda$ with $|\mu|=|\lambda|$.
This kind of structure is known for the tau function of the KP hierarchy due to Sato's theory of the KP hierarchy \cite{SN1, NT1, N1, N2, N3}.

\begin{prop}
Let $l<2L-1$, then we have $\xi_{(n_1,\cdots,n_{l})}=0$.
\end{prop}

\noindent\pf\,
We can assume $L\ge 2$.
We prove the proposition by induction on $l\le2L-2$.
We get the case of $l=1$ by Lemma 1.
Considering the case of $l=2$ in (\ref{234})  and using Lemma 1, we have
\bea
&&
\xi_{(n_1,n_2)}=\xi_{(n_1,\lambda_1,\cdots,\lambda_{2L-1})}\xi_{(n_2)}-\xi_{(n_2,\lambda_1,\cdots,\lambda_{2L-1})}\xi_{(n_1)}=0.
\non
\ena 
Thus Proposition 4 holds in the case of $l=1$ and $l=2$.

Let us consider $2\le l \le 2L-3$.
Suppose that Proposition 4 is valid for any $l'\le l$.
If $l+1$ is even, using (\ref{234}) we have
\bea
&&
\xi_{(n_1,\cdots,n_{l+1})}=\sum_{i=1}^{l+1}(-1)^{i-1}\xi_{(n_i,\lambda_1,\cdots,\lambda_{2L-1})}\xi_{(n_1,\cdots,\hat{n}_i,\cdots,n_{l+1})}.
\non
\ena
By the hypothesis of induction we have $\xi_{(n_1,\cdots,\hat{n}_i,\cdots,n_{l+1})}=0$.
Then $\xi_{(n_1,\cdots,n_{l+1})}=0$.

If $l+1$ is odd, then we have the following equation using (\ref{2301}) and the assumption of induction: 
\bea
&&
\xi_{(n_1,\cdots,n_{l+1})}=\sum_{i=1}^{2L-1}(-1)^{i-1}\xi_{(n_{l+1},\lambda_1,\cdots,\hat{\lambda}_i,\cdots,\lambda_{2L-1})}\xi_{(n_1,\cdots,n_{l},\lambda_i)}.
\label{239}
\ena
We prove $\xi_{(n_1,\cdots,n_{l},\lambda_i)}=0$.
In the same way of the proof of Proposition 2, we set $n_1=\lambda_{i_1},\cdots,n_l=\lambda_{i_l}$ in (\ref{239}) where $\{i_1,\cdots,i_l\}$ is  an arbitrary subset of $\{1,\cdots,2L-1\}$.
Then we have
\bea
&&
\xi_{(\lambda_{i_1},\cdots,\lambda_{i_l},n_{l+1})}=\sum_{i=1}^{2L-1}(-1)^{i-1}\xi_{(n_{l+1},\lambda_{1},\cdots,\hat{\lambda}_i,\cdots,\lambda_{2L-1})}\xi_{(\lambda_{i_1},\cdots,\lambda_{i_l},\lambda_i)}.
\non
\ena
Since  $l+1\le 2L-2$, the weight of $(\lambda_{i_1},\cdots,\lambda_{i_l},\lambda_i)$ is less than $|\lambda|$.
Then we obtain
\bea
&&
\xi_{(\lambda_{i_1},\cdots,\lambda_{i_l},\lambda_i)}=0.
\label{241}
\ena
Using (\ref{241}) we prove the following equation in order:
\bea
&&
\xi_{(\lambda_{i_1},\cdots,\lambda_{i_l},n_{l+1})}=0,
\non
\\
&&
\hskip1cm\vdots
\non
\\
&&
\xi_{(\lambda_{i_1},\lambda_{i_2},n_3,\cdots,n_{l+1})}=0.
\non
\ena
We can finally prove $\xi_{(\lambda_{i_1},n_2,\cdots,n_{l+1})}=0$, then the equation (\ref{239}) becomes 0.
Thus we prove $\xi_{(n_1,\cdots,n_l)}=0$.\qed

\renewcommand{\theequation}{A.\arabic{equation}}
\setcounter{equation}{0}
\appendix
\section{The neutral fermions}
A brief review on fermions is given here.
Let $\phi_n$ satisfy
\bea
&&
[\phi_m,\phi_n]_+=(-1)^m\delta_{m,-n}.
\non
\ena
In particular, we have $\phi_0^2=1/2$.
We have the properties of the vacuum state and the dual vacuum state:
\bea
&&
\phi_n|0\rangle=0,\,\,\,n<0,
\non
\\
&&
\langle0|\phi_n=0,\,\,\,n>0.
\non
\ena

Define $H_B(x)$ by
\bea
&&
H_B(x)=\frac{1}{2}\sum_{l:odd}\sum_{n\in\mathbb{Z}}(-1)^{n+1}x_l\phi_n\phi_{-n-l},
\non
\ena
and $\phi(k)$ as the generating function of $\{\phi_n\}$:
\bea
&&
\phi(k)=\sum_{n\in\mathbb{Z}}\phi_n k^n.
\non
\ena
The following the boson-fermion correspondence is valid:
\bea
&&
\langle0|\phi_0e^{H_B(x)}\phi(k)=2^{-1}X_B(k)\langle0|e^{H_B(x)},
\non
\\
&&
\langle0|e^{H_B(x)}\phi(k)=X_B(k)\langle0|\phi_0e^{H_B(x)}.
\non
\ena 

\renewcommand{\theequation}{B.\arabic{equation}}
\setcounter{equation}{0}
\section{The proof of (\ref{eq:51})}
The expansion (\ref{eq:51}) can be proved easily.
For $\alpha=(\alpha_1,\alpha_2,\cdots)$ and $\beta=(\beta_1,\beta_2,\cdots)$ we have
\bea
&&
\prod_{i,j}\frac{1+\alpha_i \beta_j}{1-\alpha_i \beta_j}=\sum_{\lambda:\text{strict}}2^{-l(\lambda)}Q^{sym}_{\lambda}(\alpha)Q^{sym}_{\lambda}(\beta)
\non
\ena
from \cite{Mac} p.255 (8.13).
The left hand side becomes
\bea
&&
\prod_{i,j}\frac{1+\alpha_i \beta_j}{1-\alpha_i \beta_j}=e^{\sum_{i,j}(\log(1+\alpha_i \beta_j)-\log(1-\alpha_i \beta_j))}
\non
\\
&&
\hskip2cm=e^{\sum_{i,j}\sum_{k=1}^{\infty}\left(\frac{(\alpha_i \beta_j)^k}{k}-\frac{(-\alpha_i \beta_j)^k}{k}\right)}
\non
\\
&&
\hskip2cm=e^{2\sum_{k:\text{odd}}k\sum_{i}\frac{\alpha_i^k}{k}\sum_{j}\frac{\beta_j^k}{k}}.
\non
\ena
Set
\bea
&&
x_k=\sum_{i}\frac{\alpha_i^k}{k},\hskip5mm y_k=\sum_{j}\frac{\beta_i^k}{k}.
\non
\ena
Then we have
\bea
&&
e^{2\sum_{k:odd}k x_k y_k}=\sum_{\lambda:\text{strict}}2^{-l(\lambda)}Q_{\lambda}(x)Q_{\lambda}(y).
\non
\ena
Replace $x_k$ by $x_k/2$ $y_k$ by $\tilde{\partial}_{y_k}=\partial_{y_k}/k$.
We apply it to $f(y)$ and set $y=0$.
Then we get
\bea
&&
f(x)=\left.\sum_{\lambda:\text{strict}}2^{-l(\lambda)}Q_{\lambda}\left(\frac{x}{2}\right)\left(Q_{\lambda}(\tilde{\partial}_y)f(y)\right)\right|_{y=0}.
\non
\ena
Setting $\xi_{\lambda}=2^{-l(\lambda)}Q_{\lambda}(\tilde{\partial}_y)f(y)|_{y=0}$, we obtain (\ref{eq:51}).

\begin{flushleft}{\Large {\bf Acknowledgments}}\end{flushleft}
I would like to thank Takashi Takebe for several comments.
I also thank Hirofumi Yamada for his interest in my work.
Finally I am deeply grateful to Atsushi Nakayashiki for much advice.
This work was supported by Japanese Association of University Woman JAUW.

\end{document}